\documentclass[journal]{IEEEtran}
\usepackage{cite}

\ifCLASSINFOpdf
   \usepackage[pdftex]{graphicx}
\else
\fi
\usepackage{amsmath}
\usepackage[linesnumbered,ruled,vlined]{algorithm2e}
\begin{document}
%
\title{A Supervised-Learning based Hour-Ahead \\ Demand Response of a Behavior-based HEMS approximating MILP Optimization}
%
%
%

\author{Huy~Truong~Dinh,~\IEEEmembership{Student Member,~IEEE,}
		Kyu-haeng Lee,~\IEEEmembership{Member,~IEEE,}
        and~Daehee~Kim,~\IEEEmembership{Member,~IEEE}
\thanks{This work was supported by the Korea Institute of Energy Technology Evaluation and Planning(KETEP) and the Ministry of Trade, Industry \& Energy(MOTIE) of the Republic of Korea (No. 20184030202130) and this work was supported by the Soonchunhyang University Research Fund.}
\thanks{The authors are with department of Future Convergence Technology, Soonchunhyang University, Asan 31538, South Korea (e-mail: tdhuy@sch.ac.kr, daeheekim@sch.ac.kr)}}

%
%

\markboth{Journal of \LaTeX\ Class Files}%
{Shell \MakeLowercase{\textit{et al.}}: Bare Demo of IEEEtran.cls for IEEE Journals}
%



\maketitle

\begin{abstract}
The demand response (DR) program of a traditional HEMS usually intervenes appliances by controlling or scheduling them to achieve multiple objectives such as minimizing energy cost and maximizing user comfort. In this study, instead of intervening appliances and changing resident behavior, our proposed strategy for hour-ahead DR firstly learns appliance use behavior of residents and then silently controls ESS and RES to minimize daily energy cost based on its knowledge. To accomplish the goal, our proposed deep neural networks (DNNs) models approximate MILP optimization by using supervised learning. The datasets for training DNNs are created from optimal outputs of a MILP solver with historical data. After training, at each time slot, these DNNs are used to control ESS and RES with real-time data of the surrounding environment. For comparison, we develop two different strategies named multi-agent reinforcement learning-based strategy, a kind of hour-ahead strategy and forecast-based MILP strategy, a kind of day-ahead strategy. For evaluation and verification, our proposed strategies are applied at three different real-world homes with real-world real-time global horizontal irradiation and real-world real-time prices. Numerical results verify that the proposed MILP-based supervised learning strategy is effective in term of daily energy cost and is the best one among three proposed strategies.
\end{abstract}


%
\IEEEpeerreviewmaketitle

\section{Introduction}
%
%
%
%
\IEEEPARstart{W}{ith} the remarkable development of renewable energy systems (RESs) and energy storage systems (ESSs), a home energy management system (HEMS) is getting more and more important in helping residents to reduce their energy consumption. Building an optimal demand response (DR) of a HEMS, which is defined as the changes in electric usage at resident side in response to changes in electricity price and surrounding environment \cite{kwac2014household}, has received much attention from researchers. Nowadays, with the amazing development of machine learning, specially deep learning techniques, many intelligent strategies for DR program, which adapt to real-time changes of environment, have been proposed.

In many previous studies (e.g. \cite{li2018real},\cite{althaher2015automated}), appliances are generally divided into two categories: non-shiftable devices and shiftable devices. For each shiftable device, residents first set up appropriate time slots in which the device should be run. With the DR program, a HEMS then tries to find a optimal day-ahead schedule for shiftable devices, satisfying all constraints of appliances and achieving multiple objectives. To find such a schedule, an explicit optimization model is built and solved. For example, in \cite{bouakkaz2021efficient}, an optimization model for minimizing energy cost is built and particle swarm optimization (PSO) algorithm is used to solve this model and find an optimal day-head schedule. In \cite{ahmad2017optimized}, an optimization model, which minimizes both energy cost and peak-to-average ratio (PAR), is built and a day-ahead schedule for appliances is found by using genetic algorithm (GA). In \cite{dinh2020home}, authors propose an optimization model in which energy selling is supported and a day-ahead schedule of appliances and selling operation is found by PSO algorithm. Beside heuristic methods, mathematical solvers are usually used to solve explicit optimization models. Authors in \cite{sou2011scheduling} build a MILP model for minimizing energy cost and use CPLEX solver to find a day-ahead schedule for appliances. In \cite{anvari2014optimal}, a MINLP solver is used to solve a multi-objective optimization model which minimizes energy cost and maximizes use comfort. Likewise, in \cite{dinh2021optimal}, a multi-objective MINLP model that jointly optimizes four objectives: energy cost, user comfort, PAR, and waiting time is built and solved by Cplex/Conopt solvers. In these aforementioned studies, constructing explicit optimization models requires detailed domain knowledge and day-ahead forecast data such as day-ahead outdoor temperature. These forecast data inevitably contain errors which degrade the performance of DR.

To overcome these challenges, learning-based approaches are introduced. For example, in \cite{ahmed2017residential}, three hidden Markov models (HMM) are built to learn the probability of each living activity and operational time of appliances. Following these models, an agent assesses running requests of appliances based consumption constraint, convenience, and grid signals. Based on an actor-critic reinforcement learning (RL), authors in \cite{bahrami2017online} develop an online load scheduling learning algorithm with real-time pricing for optimal scheduling of the controllable appliances. Multi-agent (MA) Q-learning algorithm is adopted in \cite{xu2020multi} and \cite{lu2019demand}, where the on-off operations and discrete power inputs of appliances are considered. Each agent represented for each appliance is cooperated together to minimize the electricity cost and the dissatisfaction cost.

Although RL agents do not require prior knowledge, their applicability are usually limited to domains with fully observable, low-dimensional state spaces because of their instability or even divergence \cite{mnih2015human}. Deep RL (DRL) techniques in which a deep neural network (DNN) is used to approximate an action-value function of complex, high-dimensional state spaces have been presented in many hour-ahead residential DRs. For instance, authors in \cite{valladares2019energy} propose a deep Q-learning (DQN)-based algorithm that optimizes energy consumption of heating, ventilation, and air-conditioning (HVAC) and maintains thermal comfort and air quality at the given levels. In \cite{mocanu2018line}, DQN and Deep Policy Gradient (DPG) methods are used for on-line energy scheduling electric vehicle and buildings appliances. In \cite{yu2019deep}, a deep deterministic policy gradient (DDPG) algorithm is adopted to schedule ESS and HVAC to minimize energy cost and satisfy comfortable temperature range in a smart home. For real-time scheduling of both discretely and continuously controlled appliances, in \cite{li2020real}, the trust region policy optimization (TRPO)-based algorithm, a kind of the DRL algorithm, with real-time electricity price and outdoor temperature are proposed.

The DRL techniques have achieved big success in above hour-ahead DR programs. However, in some problems, common DRL approaches sometimes are unstable or difficult to converge \cite{chen2020stabilization}. With such kinds of problems, imitation learning (IL) is used in some studies. For example, in \cite{kim2020supervised} and \cite{dinh2021milp}, authors propose optimal DRs for HVAC systems in which DRs firstly learn a mapping (or a policy) between optimal actions of a MILP solver with historical states of environment and then apply their knowledge to schedule HVAC systems with real-time data of environment. In other words, DRs try to approximate MILP optimization. The DNNs of these DRs are trained by using supervised learning technique. Their simulations show that the results of IL based method are better than those of DDPG based method. In \cite{9585298}, authors develop an IL based online power scheduling for real-time energy management of a micro-grid. In this study, their MINLP problem is firstly simplified via piece-wise linear approximation and turns into a MILP problem. A DNN is then trained to approximate MILP optimization by using data labeled optimal outputs of MILP solver with historical data. In online scheduling, this DNN is used to control their device at each time slot. Their study shows that their IL based approach outperforms proximal policy optimization (PPO) based approach.

In these aforementioned studies, DR generally intervenes operation of appliances and changes appliance use behavior of residents. In this study, we propose a supervised-learning based strategy for a hour-ahead DR of the HEMS in which DR learns resident behavior and only controls ESS and RES silently to minimize daily energy cost based on its knowledge. Residents continue to use their home devices as usual and achieve maximum comfortable lifestyle. In our study, DR imitates the behaviors of a MILP solver for hour-ahead control of ESS and RES. First, DNNs of DR are trained to learn a MILP approximation by using supervised learning technique, which is a traditional approach of IL. The dataset used for training is labeled by optimal actions of a MILP solver with historical data. After training, these DNNs are then used to control ESS and RES with current real-time data at each time slot. The main contributions of our study are as follows:
\begin{itemize}
\item We build a daily energy cost minimization problem for a smart home in the appearance of ESS, RES, and energy exchange between the smart home and other residents. Then, we reformulate the problem to another version with fewer decision variables.
\item On the basis of these formulas, we propose three strategies: MILP-based supervised learning strategy, multi-agent deep deterministic policy gradient (MADDPG)-based strategy, and forecast-based MILP strategy. First two strategies are kinds of hour-ahead strategies whereas the last strategy is a day-ahead strategy.
\item Three extensive case studies based on real-world real-time data are performed to evaluate these strategies. Numerical results show that MILP-based supervised learning strategy is effective in terms of daily energy cost when residents have a good appliance use behavior and is the best one among three proposed strategies.
\end{itemize}

The rest of this paper is organized as follows. Section II formulates our problem. In Section III, the details of three strategies are described. Case studies and simulation results are provided in Section IV. Some ideas are discussed in Section V. Finally, Section VI outlines the conclusion and future works.

\section{System Model and Problem Formulation}
\begin{figure}[]
\centering
\includegraphics[scale=0.4]{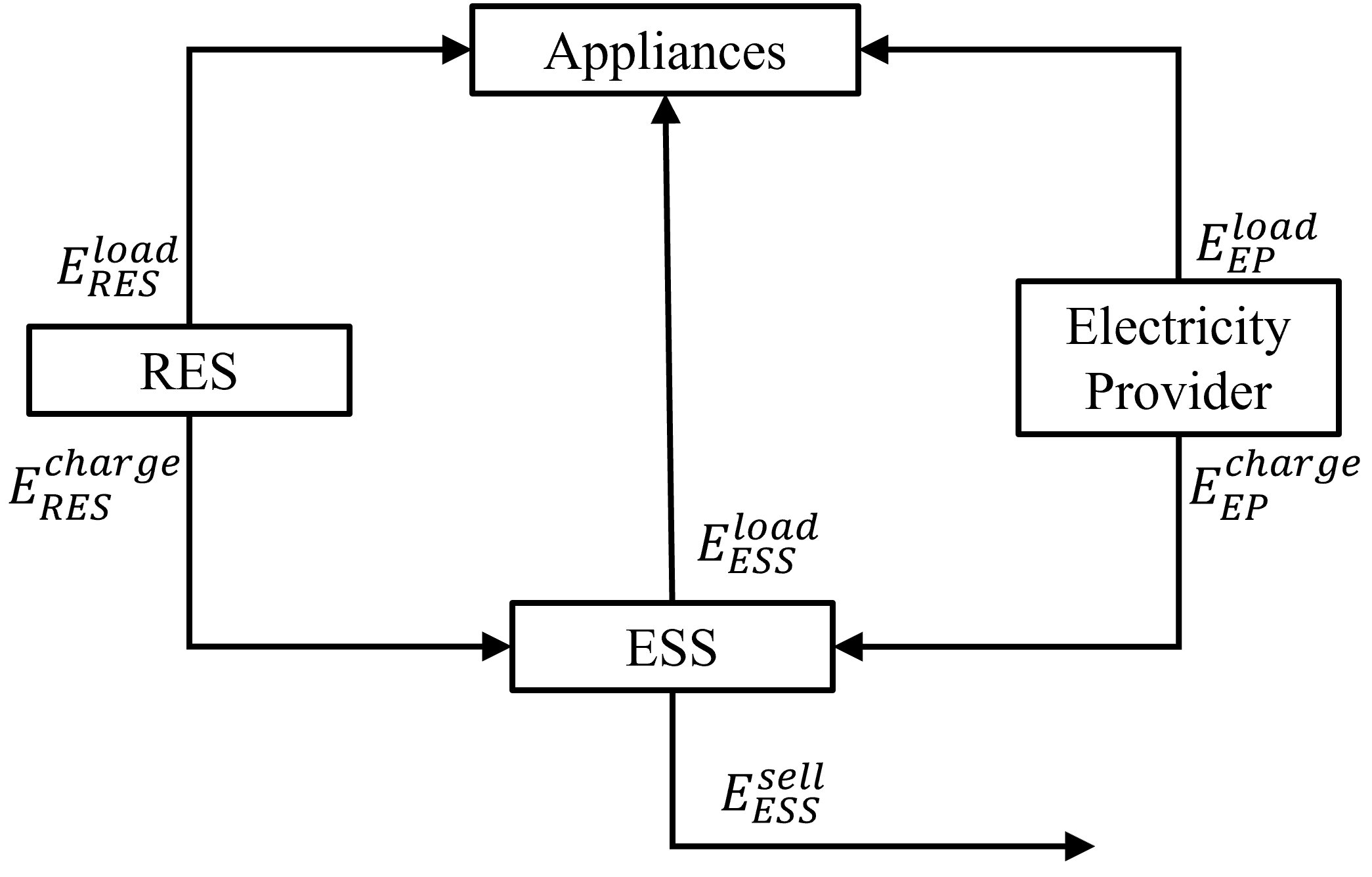}
\caption{Components and energy flows in our HEMS \cite{dinh2021optimal}.}
\label{electricity_flow}
\end{figure}

The components of a smart home considered in this study is shown in Fig. \ref{electricity_flow}, where an ESS and a PV system as RES are represented.

The electricity provider can be any outside company which can sell electricity for home. The main goals of using the ESS and the PV are to reduce energy demand from the electricity provider and allow residents to sell surplus energy to other residents. We assume that all energy to be sold comes from ESS, and RES energy is used for home load and ESS charging. In the following parts, ESS and RES models are provided and a daily energy cost minimization problem is then built for a day. We also divide a day into $T=24$ time slots, and the duration of each time slot is $\Delta t=1h$.

\subsection{ESS Model}
Let $E_{ESS}^{Level}(t)$ be the energy level of ESS after time slot $t$. As described in Fig. \ref{electricity_flow}, with $\forall t, \hspace{0.1cm}1 \leq t \leq T$, we have the following formula.
\begin{align}
E_{ESS}^{Level}(t) = E_{ESS}^{Level}(t-1) &+ \Big(E_{RES}^{charge}(t) + E_{EP}^{charge}(t)\Big) \cdot \eta^{ESS} \nonumber \\
&- \Big(E_{ESS}^{load}(t) + E_{ESS}^{sell}(t) \Big) / \eta^{ESS} 
\label{ESS_Level}
\end{align}
where $E_{ESS}^{load}(t)$ is an energy quantity used for appliances in the time slot $t$. $E_{ESS}^{sell}(t)$ is an energy quantity used to sell to the outside in the time slot $t$. $E_{RES}^{charge}(t)$ is an energy quantity stored in ESS from RES in the time slot $t$. $E_{EP}^{charge}(t)$ is an energy quantity stored in ESS from the electricity provider in the time slot $t$. $\eta^{ESS}$ is ESS efficiency. 

When using the ESS, we must satisfy the following constraints.
\begin{equation}
EL_{min} \leq E_{ESS}^{Level}(t) \leq EL_{max}
\label{ESS_Level_Constraint}
\end{equation}
\begin{equation}
E_{RES}^{charge}(t) + E_{EP}^{charge}(t) \leq Ch_{rate} \cdot \Delta t \cdot mode_{ESS}(t)
\label{ESS_Charge_Constraint}
\end{equation}
\begin{equation}
E_{ESS}^{load}(t) + E_{ESS}^{sell}(t) \leq Dh_{rate} \cdot \Delta t \cdot \big(1 - mode_{ESS}(t)\big)
\label{ESS_Discharge_Constraint}
\end{equation}
\begin{equation}
mode_{ESS}(t) = 
   \begin{cases}
   1 & \quad \text{if ESS is charged in time slot } t \\
   0 & \quad \text{if ESS is discharged in time slot } t
   \end{cases}
\label{charge_discharge_ESS_mode}
\end{equation}
where $EL_{min}$ and $EL_{max}$ are the minimum energy level and the maximum energy level of ESS. $Ch_{rate}$ and $Dh_{rate}$ are the maximum charge and discharge rate of ESS. $mode_{ESS}(t)$ is a binary variable to avoid the simultaneous ESS charging and discharging in the time slot $t$. ESS is assumed to be unable to be charged and discharged simultaneously.

Since we only consider our system during a day (no net accumulation for next day), energy level should be returned to the initial energy level $EL_0$, at the end of the day. Thus, we have
\begin{equation}
E_{ESS}^{Level}(T) = EL_0.
\label{ESS_last_level_Constraints}
\end{equation}


\subsection{RES Model}
According to \cite{ru2012storage}, output energy $E_{RES}(t)$, from a PV system in kWh in any time slot $t$ ($1 \leq t \leq T$) can be measured as 
\begin{equation}
E_{RES}(t)= GHI(t) \cdot S \cdot \eta^{RES} \cdot \Delta t.
\label{RES_power}
\end{equation}
where $GHI(t)$ is the global horizontal irradiation $(kW/m^2)$ at the location of solar panels in the time slot $t$. $S$ is the total area $(m^2)$ of solar panels and $\eta^{RES}$ is the solar conversion efficiency of the PV system.

As shown in Fig. \ref{electricity_flow}, this energy can be used for appliances and ESS charging. Thus, we have the following constrain.
\begin{equation}
E_{RES}^{load}(t) + E_{RES}^{charge}(t) \leq E_{RES}(t)
\label{RES_elements}
\end{equation}
where $E_{RES}^{load}(t)$ is an energy quantity used for appliances in time slot $t$.

It is clear that our HEMS tries to utilize RES energy as much as possible. However, if RES energy is larger than total energy demand of all appliances and ESS charging, the remaining RES energy is wasted.

\subsection{Energy Balancing}
To keep the energy balance in the smart home, the total energy demand should be equal to the total energy supply. Hence, as shown in Fig. \ref{electricity_flow}, with $\forall t, \hspace{0.1cm}1 \leq t \leq T$, we have
\begin{equation}
E_{EC}(t) = E_{EP}^{load}(t) + E_{ESS}^{load}(t) + E_{RES}^{load}(t).
\label{energy_balance}
\end{equation}
where $E_{EC}(t)$ is the energy consumption of all home appliances in a time slot $t$.

\subsection{Daily Energy Cost Minimization Problem}
We assume that the energy from RES and ESS is complimentary and selling real-time price $P_{sell}(t)$, is related to buying real-time price $P_{realtime}(t)$, (e.g., $P_{sell}(t)= \alpha_p \cdot P_{realtime}(t)$ with $\alpha_p$ is a constant and $\alpha_p \leq 1$). Then, daily energy cost minimization problem can be formulated as
\begin{align}
min \displaystyle\sum_{t=1}^{T}C(t)= min \displaystyle\sum_{t=1}^{T} \Bigg(\Big(E_{EP}^{load}&(t) + E_{EP}^{charge}(t)\Big)\cdot P_{realtime}(t) \nonumber \\
&- E_{ESS}^{sell}(t) \cdot \alpha_p \cdot P_{realtime}(t)\Bigg)
\label{objective_cost_function_1}
\end{align}

Combining with (\ref{energy_balance}), our problem in (\ref{objective_cost_function_1}) can be turned into
\begin{align}
min \displaystyle\sum_{t=1}^{T}C(t)= min \displaystyle\sum_{t=1}^{T} &\Big(E_{EC}(t) - E_{RES}^{load}(t)+ E_{EP}^{charge}(t) \nonumber \\
&- E_{ESS}^{load}(t) - \alpha_p \cdot E_{ESS}^{sell}(t) \Big) \cdot P_{realtime}(t)
\label{objective_cost_function_2}
\end{align}
\begin{equation}
s.t. (\ref{ESS_Level}) - (\ref{energy_balance}) \nonumber
\end{equation}

It is clear that if we know energy consumption $E_{EC}(t)$, real-time irradiation $GHI(t)$, and real-time price $P_{realtime}(t)$ at every time slot $t$ of a day, our problem in (\ref{objective_cost_function_2}) is a MILP problem. Hence, by using MILP solvers, we can easily find optimal values of variables $E_{RES}^{load}(t)$, $E_{EP}^{charge}(t)$, $E_{ESS}^{load}(t)$, and $E_{ESS}^{sell}(t)$ at every time slot $t$. These optimal values are optimal energy of ESS and RES which should be used at each time slot to achieve optimal daily energy cost. Unfortunately, we only know these information at the end of the day. It means that MILP solvers are only useful at the end of the day and it is too late to control them. Hence, to overcome this problem and utilize powerful MILP solvers, we propose MILP-based supervised learning strategies in the next section.

\subsection{A Different Version of Daily Energy Cost Minimization Problem}
Because ESS cannot be charged or discharged at the same time slot,  we define a new float variable  $E_{ESS}^{CD}(t)$ which refers to an energy quantity stored in the ESS in a time slot $t$ if $E_{ESS}^{CD}(t) \geq 0$ and refers to an energy quantity which is drawn from ESS in a time slot $t$ if $E_{ESS}^{CD}(t) < 0$. It means that
\begin{equation}
E_{ESS}^{CD}(t) = 
   \begin{cases}
   E_{RES}^{charge}(t) + E_{EP}^{charge}(t) & \quad  E_{ESS}^{CD}(t) \geq 0 \\
   -\Big(E_{ESS}^{load}(t) + E_{ESS}^{sell}(t) \Big) & \quad  E_{ESS}^{CD}(t) < 0 \\
   \end{cases}
\label{E_CD_1}
\end{equation}

Hence, we have
\begin{equation}
-Dh_{rate} \cdot \Delta t \leq E_{ESS}^{CD}(t) \leq Ch_{rate} \cdot \Delta t
\label{E_CD_2}
\end{equation}
\begin{equation}
E_{ESS}^{level}(t) = 
   \begin{cases}
    E_{ESS}^{level}(t-1) + E_{ESS}^{CD}(t) \cdot \eta^{ESS}& \mkern5mu  E_{ESS}^{CD}(t) \geq 0 \\
    E_{ESS}^{level}(t-1) + E_{ESS}^{CD}(t) / \eta^{ESS} & \mkern5mu  E_{ESS}^{CD}(t) < 0 \\
   \end{cases}
\label{E_CD_3}
\end{equation}

When ESS is in charge mode ($E_{ESS}^{CD}(t) \geq 0$) in a time slot $t$, we have $E_{ESS}^{load}(t) = E_{ESS}^{sell}(t) = 0$. Hence, combining with (\ref{E_CD_1}), energy cost in this time slot can be calculated as follows.
\begin{align}
C(t) = \Big(E_{EC}(t) - E_{RES}^{load}&(t) + E_{ESS}^{CD}(t) \nonumber \\
&- E_{RES}^{charge}(t) \Big) \cdot P_{realtime}(t) \nonumber \\
\Rightarrow C(t)= \Big(E_{EC}(t) - E_{RES}^{load}&(t) + E_{ESS}^{CD}(t) \nonumber \\
&- RC(t) \Big) \cdot P_{realtime}(t)			
\label{objective_cost_function_3}
\end{align}
where variable $RC(t)$, which depends on $E_{RES}^{load}(t)$ and $E_{ESS}^{CD}(t)$, is calculated as
\begin{equation}
RC(t) = min\{E_{RES}(t) - E_{RES}^{load}(t),E_{ESS}^{CD}(t)\}.
\label{RC_constraint}
\end{equation}

When ESS is in discharge mode ($E_{ESS}^{CD}(t) < 0$) in a time slot $t$, we have $E_{RES}^{charge}(t) = E_{EP}^{charge}(t) = 0$. Hence, energy cost in this time slot can be calculated as follows.
\begin{equation}
C(t) = E_{EP}^{load}(t) \cdot P_{realtime}(t) - E_{ESS}^{sell}(t) \cdot \alpha_{p} \cdot P_{realtime}(t) \nonumber
\end{equation}

Because $P_{sell}(t) \leq P_{realtime}(t)$, we only sell energy to outside if total energy supply is larger than energy consumption of home all appliances in this time slot. Hence, we have
\begin{equation}
C(t) = 
   \begin{cases}
    \Big(E_{EC}(t) - E_{RES}^{load}(t) + E_{ESS}^{CD}(t)\Big) \cdot P_{realtime}(t) \\  & \mkern-288mu \textit{if } E_{EC}(t) \geq E_{RES}^{load}(t) - E_{ESS}^{CD}(t) \\
    \Big(E_{EC}(t) - E_{RES}^{load}(t) + E_{ESS}^{CD}(t)\Big) \cdot \alpha_p \cdot P_{realtime}(t) \\  & \mkern-288mu \textit{if } E_{EC}(t) < E_{RES}^{load}(t) - E_{ESS}^{CD}(t) \\
   \end{cases}
\end{equation}

It is worth noting that in discharge mode, all RES energy is only used for appliances. Let $RL(t)= min\{E_{EC}(t) - E_{RES}(t),0\}$, we have
\begin{equation}
C(t) = 
   \begin{cases}
    \Big(RL(t) + E_{ESS}^{CD}(t)\Big) \cdot P_{realtime}(t)  \\&\mkern-144mu \textit{if } RL(t) \geq -E_{ESS}^{CD}(t) \\
    \Big(RL(t) + E_{ESS}^{CD}(t)\Big) \cdot \alpha_p \cdot P_{realtime}(t) \\ &\mkern-144mu \textit{if } RL(t) < - E_{ESS}^{CD}(t) \\
   \end{cases}
\label{objective_cost_function_4}   
\end{equation}

In summary, our daily energy cost minimization problem can be formulated as follows.
\begin{equation}
min \displaystyle\sum_{t=1}^{T}C(t)
\label{objective_cost_function_5}
\end{equation}
\begin{equation}
s.t. (\ref{ESS_Level_Constraint}), (\ref{ESS_last_level_Constraints}), (\ref{RES_power}), (\ref{RES_elements}), (\ref{E_CD_2}), (\ref{E_CD_3}), (\ref{RC_constraint}) \nonumber
\end{equation}
where $C(t)$ is calculated as in (\ref{objective_cost_function_3}) if $E_{ESS}^{CD} \geq 0$ and is calculated as in (\ref{objective_cost_function_4}) if $E_{ESS}^{CD} < 0$.

Although this version of our minimization problem is not a MILP problem, it only depends on two variables: $E_{ESS}^{CD}(t)$ and $E_{RES}^{load}(t)$. To solve this version, we propose multi-agent DRL-based strategy in which an agent controls $E_{ESS}^{CD}(t)$ of ESS and another agent controls $E_{RES}^{load}(t)$ of RES to minimize daily energy cost.

\section{Three optimal strategies}

In this section, we propose three strategies: MILP-based supervised learning, MADDPG-based strategy and forecast-based MILP strategy. First two strategies are kinds of hour-ahead strategies while the last strategy is a kind of day-ahead strategy.

\subsection{MILP-based supervised learning strategy}
\begin{figure*}[ht]
\centering
\includegraphics[scale=0.45]{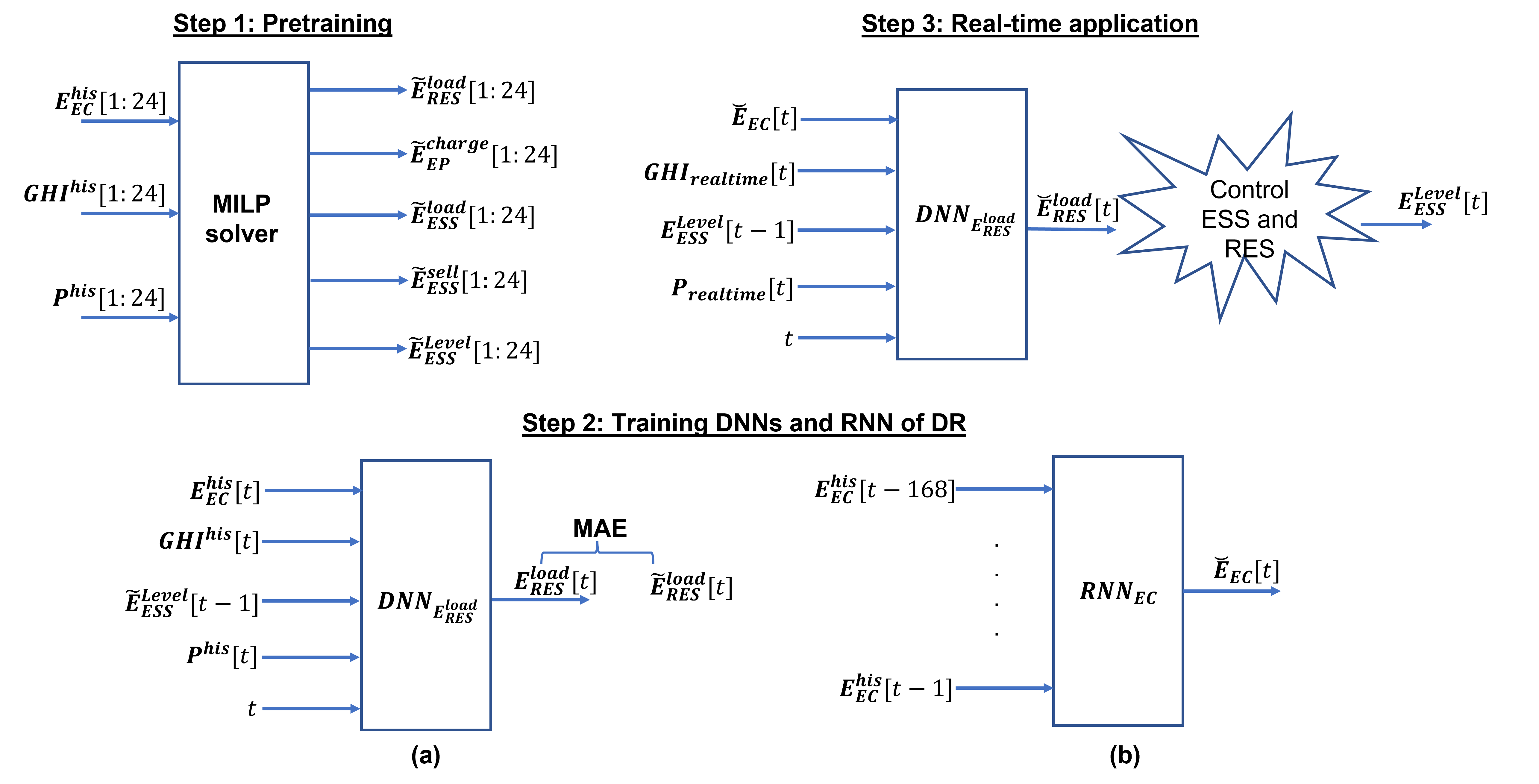}
\caption{Overall framework of MILP-based supervised learning: (a) training DNNs, (b) training an RNN to predict energy consumption for next $1$ time slot.}
\label{framework}
\end{figure*}

In this strategy, at the beginning, our problem in (\ref{objective_cost_function_2}) is solved by a MILP solver with necessary data from the surrounding environment of historical days. The results of this process are optimal values of variables of ESS and RES we should use to control at each time slot of these historical days. From these optimal values, datasets for these variables are built and then used to train DNNs which will learn an approximation of a MILP solver. After training, DNNs are used to control them at each time slot with current real-time data. However, these DNNs also need the forecast value of energy consumption of all appliances at each time slot as their input data. Hence, our HEMS needs to learn appliance use behavior of residents to predict the energy consumption for next $1$ time slot. Fig. \ref{framework} shows the overall framework of MILP-based supervised learning in detail.

In step $1$, the historical energy consumption $E_{EC}^{his}[1:24]$, the historical irradiation $GHI^{his}[1:24]$, and historical prices $P^{his}[1:24]$ of a historical day are input data for the MILP solver. The output of the MILP solver are the optimal energy which should be used to control ESS and RES at $24$ time slots of this historical day: $\widetilde{E}_{RES}^{load}[1:24]$, $\widetilde{E}_{EP}^{charge}[1:24]$, $\widetilde{E}_{ESS}^{load}[1:24]$, $\widetilde{E}_{ESS}^{sell}[1:24]$. Moreover, the MILP solver also gives us optimal energy levels $\widetilde{E}_{ESS}^{Level}[1:24]$ of ESS after $24$ time slots of this day. A training dataset $D_{E_{RES}^{load}}$ for variable $E_{RES}^{load}(t)$, is then structured with formatting as follows:
\begin{align}
input&= \begin{aligned}[t]
     \Big[E_{EC}^{his}[t], GHI^{his}[t], \widetilde{E}_{ESS}^{Level}[t-1], P^{his}[t], t \Big] \nonumber
    \end{aligned}\\
label&=\begin{aligned}[t]
&\Big[\widetilde{E}_{RES}^{load}[t] \Big] \nonumber
       \end{aligned}
\end{align}

By changing the content of label into $\widetilde{E}_{EP}^{charge}[t]$, we will have new training dataset $D_{E_{EP}^{charge}}$ for variable $E_{EP}^{charge}(t)$. With the similar changing, we also have new training datasets $D_{E_{ESS}^{load}}$ and $D_{E_{ESS}^{sell}}$ for variables $E_{ESS}^{load}(t)$ and $E_{ESS}^{sell}(t)$ respectively.

In step $2$, four datasets $D_{E_{RES}^{load}}$, $D_{E_{EP}^{charge}}$, $D_{E_{ESS}^{load}}$ and $D_{E_{ESS}^{sell}}$ are used to train four neural networks $DNN_{E_{RES}^{load}}$, $DNN_{E_{EP}^{charge}}$, $DNN_{E_{ESS}^{load}}$, and $DNN_{E_{ESS}^{sell}}$ of our HEMS, respectively. It is worth noting that in Fig. \ref{framework}, only $DNN_{E_{RES}^{load}}$ is shown. We also need to train a recurrent neural network (RNN) to predict amount of energy consumption of the home in next $1$ time slot. In this study, we use RNN with GRU cell $RNN_{EC}$, which includes $2$ layers and time step is $168$.

Finally, in step $3$, our HEMS uses four trained DNNs and $RNN_{EC}$ to control ESS and RES in $1$ time slot. At the beginning of each time slot $t$ in a day, firstly, $RNN_{EC}$ is used to compute forecast value $\breve{E}_{EC}[t]$ of the energy consumption $E_{EC}(t)$ in this time slot. Then $[\breve{E}_{EC}[t], GHI_{realtime}[t], E_{ESS}^{Level}[t-1], P_{realtime}[t], t]$ are input data for $DNN_{E_{RES}^{load}}$, and its output is the forecast value $\breve{E}_{RES}^{load}[t]$ of variable $E_{RES}^{load}(t)$. Likewise, these input data also are used as input data for the remaining DNNs and forecast values of the remaining variables are achieved. After achieving four forecast values of four variables, they are used to control ESS and RES in this time slot. It is worth noting that, in this step, all input data are current real-time values except for $\breve{E}_{EC}[t]$. Clearly, the efficiency of this strategy depends on the accuracy of energy consumption prediction in next $1$ time slot (the performance of $RNN_{EC}$) which is shown in Section IV. More detailed MILP-based supervised learning strategy can be found in Algorithm \ref{SL_algorithm}. 
\begin{algorithm}
  \caption{MILP-based supervised learning strategy}
  \label{SL_algorithm}
  \SetKwInOut{Input}{Input}
  \SetKwInOut{Output}{Output}
  \SetKwComment{Comment}{/*}{*/}
  \emph{\textbf{Step 1: Create training datasets from historical days}}\;
  \Input{Number of historical days $hd$, historical energy consumption $E_{EC}^{his}[hd][1:24]$, historical irradiation $GHI^{his}[hd][1:24]$ and historical prices $P^{his}[hd][1:24]$}
  \Output{Datasets for training the DNNs of variables}
  \For{$i \leftarrow 1$ \KwTo $hd$}{
  	Solve (\ref{objective_cost_function_2}) with $E_{EC}^{his}[i][1:24]$, $GHI^{his}[i][1:24]$, and $P^{his}[i][1:24]$\;
  	$=> \widetilde{E}_{RES}^{load}[1:24]$, $\widetilde{E}_{EP}^{charge}[1:24]$, $\widetilde{E}_{ESS}^{load}[1:24]$, $\widetilde{E}_{ESS}^{sell}[1:24]$, $\widetilde{E}_{ESS}^{Level}[1:24]$\;
  	\For{$t \leftarrow 1$ \KwTo $24$}{
  	   $input.append(\Big[E_{EC}^{his}[t],GHI^{his}[t], \widetilde{E}_{ESS}^{Level}[t-1], P^{his}[t],t \Big])$\;
  	   $label_{E_{RES}^{load}}.append(\widetilde{E}_{RES}^{load}[t])$\;
  	   $label_{E_{EP}^{charge}}.append(\widetilde{E}_{EP}^{charge}[t])$\;
  	   $label_{E_{ESS}^{load}}.append(\widetilde{E}_{ESS}^{load}[t])$\;
  	   $label_{E_{ESS}^{sell}}.append(\widetilde{E}_{ESS}^{sell}[t])$\;
  	}
  }
  \emph{\textbf{Step 2: Train DNNs and RNN}}\;
  Train a $DNN_{E_{RES}^{load}}$ with $\{input, label_{E_{RES}^{load}}\}$\;
  Train a $DNN_{E_{EP}^{charge}}$ with $\{input, label_{E_{EP}^{charge}}\}$\;  
  Train a $DNN_{E_{ESS}^{load}}$ with $\{input, label_{E_{ESS}^{load}}\}$\;
  Train a $DNN_{E_{ESS}^{sell}}$ with $\{input, label_{E_{ESS}^{sell}}\}$\;
  Train an $RNN_{EC}$ with $E_{EC}^{his}[hd][1:24]$\;
  \emph{\textbf{Step 3: Use DNNs and RNN during a day}}\;
  \For{$t \leftarrow 1$ \KwTo $24$}{ 
  	Measure real-time irradiation $GHI_{realtime}[t]$ \;
  	Calculate $E_{RES}[t]$ as in (\ref{RES_power})\;
  	$\breve{E}_{EC}[t] = RNN_{EC}(E_{EC}[t-168:t-1])$\;
  	$input=[\breve{E}_{EC}[t], GHI_{realtime}[t], E_{ESS}^{Level}[t-1], P_{realtime}[t], t]$\;
  	\emph{/*Get forecast values of variables */}\;
  	$\breve{E}_{RES}^{load}[t]= DNN_{E_{RES}^{load}}(input)$\;
    $\breve{E}_{RES}^{charge}[t] = E_{RES}[t] - \breve{E}_{RES}^{load}[t]$\;
  	$\breve{E}_{EP}^{charge}[t]= DNN_{E_{EP}^{charge}}(input)$\;
  	$\breve{E}_{ESS}^{load}[t]= DNN_{E_{ESS}^{load}}(input)$\;
  	$\breve{E}_{ESS}^{sell}[t]= DNN_{E_{ESS}^{sell}}(input)$\;
  	\emph{/*Calculate forecast energy quantity for discharging and charging */}\;
  	$\breve{EC} = \breve{E}_{RES}^{charge}[t] + \breve{E}_{EP}^{charge}[t]$\;
  	$\breve{ED} = \breve{E}_{ESS}^{load}[t] + \breve{E}_{ESS}^{sell}[t]$\;
  	\uIf(\tcp*[f]{charge mode}) {$\breve{EC} \geq \breve{ED}$}
  	{
  	Use $\breve{E}_{RES}^{load}[t], \breve{E}_{RES}^{charge}[t]$, and $\breve{E}_{EP}^{charge}[t]$ to control PV system and ESS in this time slot\;
  	}
  	\Else(\tcp*[f]{discharge mode}){
  	Discharge ESS with an energy quantity, $\breve{ED}$, for appliances. If this energy quantity is larger than real energy consumption of appliances, remaining energy is used for selling. All energy from PV system is used for appliances in this time slot\;
  	}
  	Calculate $E_{ESS}^{Level}[t]$ as in (\ref{E_CD_3})\;
  	}
\end{algorithm} 

\subsection{MADDPG-based strategy}
Our minimization problem defined in (\ref{objective_cost_function_5}) can be solved by using advanced MA-DRL algorithms. The key components of MA-DRL environment in our problem can be designed as follows:
\begin{itemize}
\item State: The environment state $s_t$ includes $5$ kinds of information at the beginning of each time slot $t$: the energy consumption $E_{EC}[t]$, the real-time irradiation $GHI_{realtime}[t]$, the current energy level of ESS $E_{ESS}^{Level}[t-1]$, the real-time price $P_{realtime}[t]$, and time slot $t$. For brevity, we denote the state by $s_t=[E_{EC}[t], GHI_{realtime}[t], E_{ESS}^{Level}[t-1], P_{realtime}[t], t]$.
\item Agents and actions: our DR includes two agents $A_{ESS}$ and $A_{RES}$ whose goals are to decide optimal values of variables $E_{ESS}^{CD}(t)$ and $E_{RES}^{load}(t)$ which should be used to control ESS and RES in the time slot $t$, respectively. Hence, the action of $A_{ESS}$ is defined as $a_{ESS}[t] = E_{ESS}^{CD}[t]$ whereas the action of $A_{RES}$ is defined as $a_{RES}[t] = E_{RES}^{load}[t]$.
\item Reward: At the beginning of each time slot $t$, when two agents execute their actions, the transition of the environment state is triggered from $s_t$ to $s_{t+1}$ and two agents receive a same reward $R_t$. Our objective is to minimize the daily energy cost. Hence, $R_t$ can be defined as $R_t=-C(t)$ with $C(t)$ is calculated as in (\ref{objective_cost_function_3}) if $a_{ESS}[t] \geq 0$ and is calculated as in (\ref{objective_cost_function_4}) if $a_{ESS}[t] < 0$.
\end{itemize}

\begin{figure}[ht]
\centering
\includegraphics[scale=0.4]{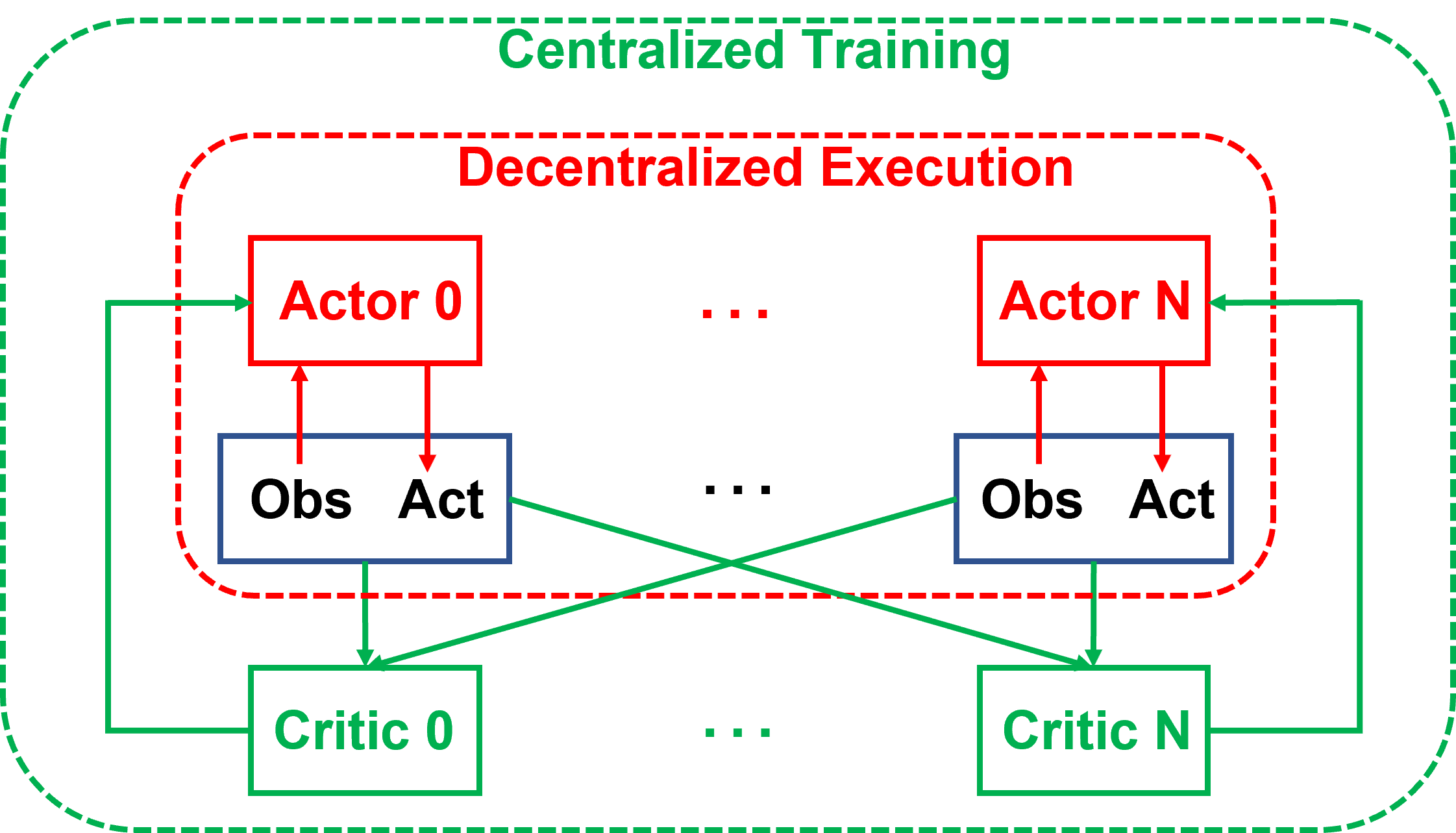}
\caption{MADDPG architecture \cite{farag2020multi}.}
\label{MADDPG}
\end{figure}

Because the action space of both agents is continuous, we propose a strategy based on MADDPG algorithm \cite{lowe2017multi} whose architecture is shown in Fig. \ref{MADDPG}.

MADDPG algorithm is an extended version of DDPG algorithm \cite{lillicrap2015continuous} applied to a MA environment and is reported to defeat other DRL algorithms like DQN, Actor-Critic, TRPO \cite{lowe2017multi}. This algorithm includes two phases: centralized training for critic updates and decentralized execution for actions. Each agent has its own continuous action space and observation space. In centralized training, each agent, firstly, try to collect information from other agents and then calculate and update its own critic network based on these joint information. As the critic network learns the joint action-value function $Q(s,a)$ over time, the deterministic policy gradient is also calculated and sent to the actor network to help update the parameters of the actor network, similar to DDPG algorithm. The most important thing to notice that even though the critic network needs joint information, the actor network can only use its own observation space to make a decision (decentralized execution). Detailed MADDPG algorithm can be found in \cite{lowe2017multi}. The brief explanation of MADDPG-based strategy is as follows.

In the step $1$, the agents $A_{ESS}$ and $A_{RES}$ are trained by using historical energy consumption $E_{EC}^{his}[1:24]$, historical irradiation $GHI^{his}[1:24]$, and historical prices $P^{his}[1:24]$. Similar to supervised learning strategy, an RNN $RNN_{EC}$, is also trained to predict amount of energy consumption of the home in next $1$ time slot.

After training all agents, in step $2$, at the beginning of each time slot $t$ of a day, the $RNN_{EC}$ is first used to compute forecast value $\breve{E}_{EC}[t]$ of the energy consumption $E_{EC}(t)$ in this time slot. Then $[\breve{E}_{EC}[t], GHI_{realtime}[t], E_{ESS}^{Level}[t-1], P_{realtime}[t], t]$ are input data for actor networks of two agents $A_{ESS}$ and $A_{RES}$. The outputs are forecast values of variables $E_{ESS}^{CD}(t)$ and $E_{RES}^{load}(t)$ which should be used to control ESS and RES. The detailed MADDPG-based strategy is shown in Algorithm \ref{MADDPG_algorithm}.
\begin{algorithm}
  \caption{MADDPG-based strategy}
  \label{MADDPG_algorithm} 
  \SetKwInOut{Input}{Input}
  \SetKwInOut{Output}{Output}
  \SetKwComment{Comment}{/*}{*/}   
  	\emph{\textbf{Step 1: Train an agent of RES $A_{RES}$, and an agent of ESS $A_{ESS}$, from historical days}}\;
   	\Input{Number of historical days $hd$, historical energy consumption $E_{EC}^{his}[hd][1:24]$, historical irradiation $GHI^{his}[hd][1:24]$ and historical prices $P^{his}[hd][1:24]$}
  	\Output{The trained $A_{RES}$ and $A_{ESS}$}
	Use MADDPG algorithm to train $A_{RES}$ for action $E_{RES}^{load}$ and $A_{ESS}$ for action $E_{ESS}^{CD}$ with $E_{EC}^{his}[hd][1:24]$, $GHI^{his}[hd][1:24]$, and $P^{his}[hd][1:24]$ \;
	Train an $RNN_{EC}$ with $E_{EC}^{his}[hd][1:24]$\;
	\BlankLine
	\emph{\textbf{Step 2: Use the agent of RES and the agent of ESS during a day}}\;
	\For{$t \leftarrow 1$ \KwTo $24$}{
	  	Measure real-time irradiation $GHI_{realtime}[t]$\;
	  	Calculate $E_{RES}[t]$ as in (\ref{RES_power})\;
  		\emph{/* Get forecast values of energy consumption */}\;
  		$\breve{E}_{EC}[t] = RNN_{EC}(E_{EC}[t-168:t-1])$\;
  		$s_t=[\breve{E}_{EC}[t], GHI_{realtime}[t], E_{ESS}^{Level}[t-1], P_{realtime}[t], t]$\;
  		\emph{/* Get forecast value of variables */}\;
  		$\breve{E}_{RES}^{load}[t]= A_{RES}(s_t)$\;
  		$\breve{E}_{ESS}^{CD}[t]= A_{ESS}(s_t)$\;
  		\uIf(\tcp*[f]{charge mode}) {$\breve{E}_{ESS}^{CD}[t] >= 0$}{ 
		    $\breve{E}_{RES}^{charge}[t] = E_{RES}[t] - \breve{E}_{RES}^{load}[t]$\;
  			$\breve{E}_{EP}^{charge}[t] = \breve{E}_{ESS}^{CD}[t] - \breve{E}_{RES}^{charge}[t]$\;
  			Use $\breve{E}_{RES}^{load}[t], \breve{E}_{RES}^{charge}[t]$, and $\breve{E}_{EP}^{charge}[t]$ to control PV system and ESS in this time slot\;
  		}
	  	\Else(\tcp*[f]{discharge mode}){
			Discharge ESS with an energy quantity, $-\breve{E}_{ESS}^{CD}[t]$, for appliances. If this energy quantity is larger than real energy consumption of appliances, remaining energy is used for selling. All energy from PV system is used for appliances in this time slot\;  			
	  	}
    	Calculate $E_{ESS}^{Level}[t]$ as in (\ref{E_CD_3})\;
  	}
\end{algorithm}

\subsection{Forecast-based MILP strategy}
The main prerequisite of using the MILP solver is that all values of energy consumption $E_{EC}(t)$, real-time irradiation $GHI_{realtime}(t)$, and real-time price $P_{realtime}(t)$ need to be known at every time slot of a day. However, we usually do not have this information until the end of the day. Hence, to overcome this problem, we propose a forecast-based MILP strategy in which all needed data for the MILP solver are forecast at the beginning of the day. The brief explanation of this strategy is as follows.

In step $1$, RNNs for prediction of energy consumption, real-time irradiation, and real-time prices for next $24$ time slots are trained.

In step $2$, at the beginning of a day, we achieve forecast values of energy consumption, real-time irradiation, and real-time prices for $24$ time slots of a day by using these RNNs. These forecast values are then used to solve (\ref{objective_cost_function_2}) by using a MILP solver. The output of the MILP solver are forecast values of all variables of ESS and RES in our problem for next $24$ time slots and they will be used to control ESS and RES at each time slot during this day. The detailed forecast-based MILP strategy is shown in Algorithm \ref{forecast_algorithm}.
\begin{algorithm}
  \caption{Forecast-based MILP strategy}
  \label{forecast_algorithm}
  \SetKwInOut{Input}{Input}
  \SetKwInOut{Output}{Output}
  \SetKwComment{Comment}{/*}{*/}     
  	\emph{\textbf{Step 1: Train RNNs of energy consumption, real-time irradiation, and real-time prices}}\;
  	Train $RNN_{EC}^{24}$ with $E_{EC}^{his}[hd][1:24]$ for next 24-hour forecast energy consumption \;
  	Train $RNN_{GHI}^{24}$ with $GHI^{his}[hd][1:24]$ for next 24-hour forecast real-time irradiation\;
  	Train $RNN_{price}^{24}$ with $P^{his}[hd][1:24]$ for next 24-hour forecast real-time prices\;
	\BlankLine
	\emph{\textbf{Step 2: Use RNNs for a day}}\;
  	\emph{ /*Get 24-hour-forecast values of energy consumption, irradiation, and price*/}\;
	$\breve{E}_{EC}^{24}[1:24] = RNN_{EC}^{24}(E_{EC}[t-168:t-1])$\;
	$\breve{GHI}_{realtime}^{24}[1:24] = RNN_{GHI}^{24}(GHI[t-168:t-1])$\;
	$\breve{P}_{realtime}^{24}[1:24] = RNN_{price}^{24}(P[t-168:t-1])$\;
	Solve (\ref{objective_cost_function_2}) with $\breve{E}_{EC}^{24}[1:24]$, $\breve{GHI}_{realtime}^{24}[1:24]$, and $\breve{P}_{realtime}^{24}[1:24]$\;
	$=>\widetilde{E}_{RES}^{load}[1:24]$, $\widetilde{E}_{RES}^{charge}[1:24]$, $\widetilde{E}_{EP}^{charge}[1:24]$, $\widetilde{E}_{ESS}^{load}[1:24]$, $\widetilde{E}_{ESS}^{sell}[1:24]$\;
	\BlankLine
	\For{$t \leftarrow 1$ \KwTo $24$}{
	  	\emph{/* Calculate forecast energy quantity for discharging and charging */}\;
	  	$\breve{EC} = \breve{E}_{RES}^{charge}[t] + \breve{E}_{EP}^{charge}[t]$\;
	  	$\breve{ED} = \breve{E}_{ESS}^{load}[t] + \breve{E}_{ESS}^{sell}[t]$\;
		\uIf(\tcp*[f]{charge mode}) {$\breve{EC} \geq \breve{ED} $}{
	  		Use $\breve{E}_{RES}^{load}[t], \breve{E}_{RES}^{charge}[t]$, and $\breve{E}_{EP}^{charge}[t]$ to control PV system and ESS in this time slot\;
	  	}
    	\Else(\tcp*[f]{discharge mode}){
			Discharge ESS with an energy quantity, $\breve{ED}$, for appliances. If this energy quantity is larger than real energy consumption of appliances, remaining energy is used for selling. All energy from PV system is used for appliances in this time slot\;  			
	  	}   
  	}
\end{algorithm}

\section{Case Studies and Simulation Results}
In this section, we describe the simulation setup and different case studies to which our proposed strategies are applied. The performance of our proposed strategies is evaluated through numerical simulation results under these case studies. For comparison, the efficiency of our proposed strategies are firstly calculated based on MILP results which are only achieved at the end of a day when we already knew full information of that day. From these results, the performance of our proposed strategies is compared together.

\subsection{Simulation Setup}
The performance of our proposed strategies depends on the accuracy of energy consumption prediction in next $1$ time slot. In other words, the performance of our strategies depends on behaviors in which residents use their appliances. Hence, for evaluation and comparison, the our proposed strategies are applied to three different real-world homes which are classified based on the resident behavior: $stable$, $fluctuating$, and $chaos$. These homes are located at London, UK and their datasets are extracted from ``Energy Consumption Data in London Households'' dataset, a real-world biggest dataset of UK Power Network from Jan 2012 to Feb 2014 \cite{household}.

\begin{figure}[ht]
\centering
\includegraphics[scale=0.035]{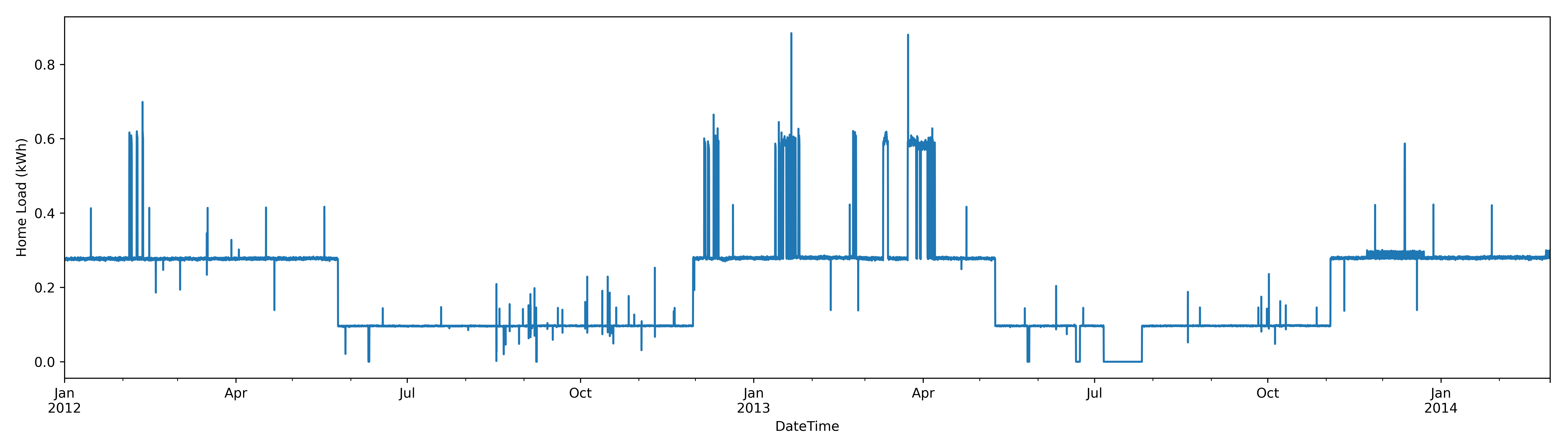}
\caption{Hourly energy consumption of a stable home from Jan 2012 to Feb 2014.}
\label{stable}
\end{figure}

The stable home describes a home in which resident behavior almost does not change in using appliances day by day. Fig. \ref{stable} shows the historical dataset of hourly energy consumption of this home. After training $RNN_{EC}$ with this dataset, the 1-hour-forecast value predicted by $RNN_{EC}$ is almost the same as the real value in the testing set. Average MAPE of testing set is only $0.6\%$.

\begin{figure}[ht]
\centering
\includegraphics[scale=0.035]{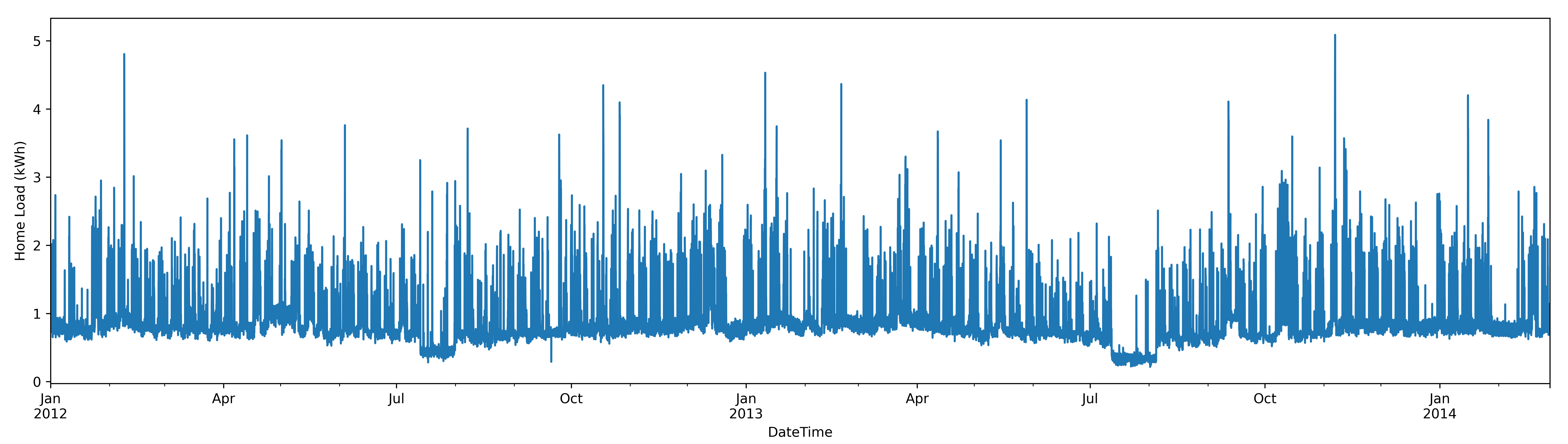}
\caption{Hourly energy consumption of a fluctuating home from Jan 2012 to Feb 2014.}
\label{fluctuating}
\end{figure}

The fluctuating home describes a home in which resident behavior changes slightly. Fig. \ref{fluctuating} shows the historical dataset of hourly energy consumption of this home. After training $RNN_{EC}$ with this dataset, the 1-hour-forecast value predicted by $RNN_{EC}$ is a little different from the real value in the testing set. Average MAPE of testing set is $10.9\%$.

\begin{figure}[ht]
\centering
\includegraphics[scale=0.035]{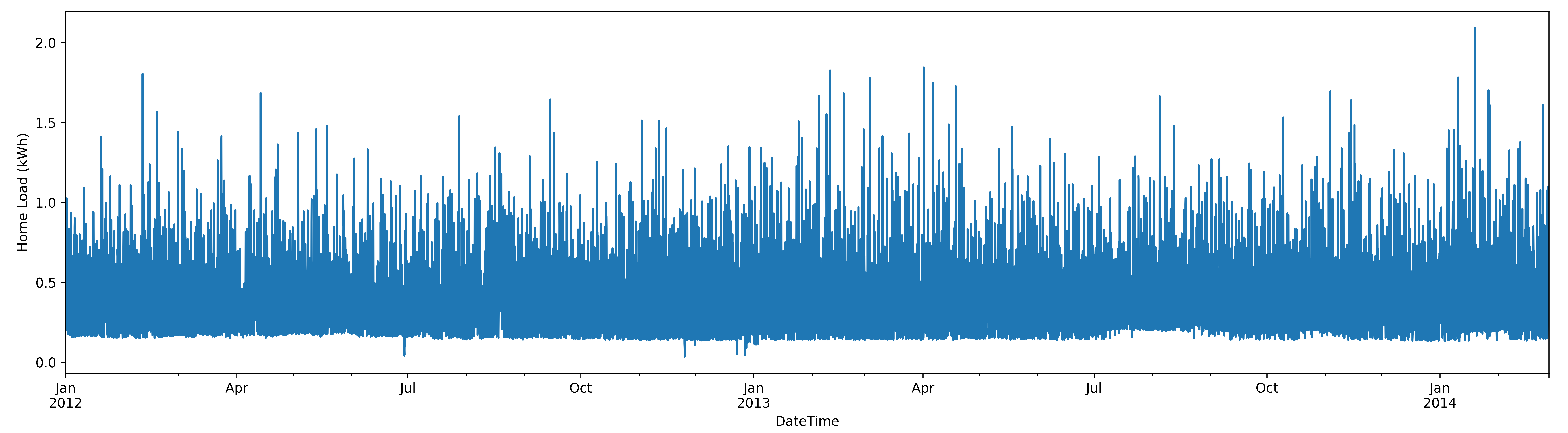}
\caption{Hourly energy consumption of a chaos home from Jan 2012 to Feb 2014.}
\label{chaos}
\end{figure}

The chaos home describes a home in which resident behavior changes a lot. Fig. \ref{chaos} shows the historical dataset of hourly energy consumption of this home. After training $RNN_{EC}$ with this dataset, the 1-hour-forecast value predicted by $RNN_{EC}$ is very different from the real value in the testing set. Average MAPE of testing set is $21.8\%$.

\begin{figure}[ht]
\centering
\includegraphics[scale=0.035]{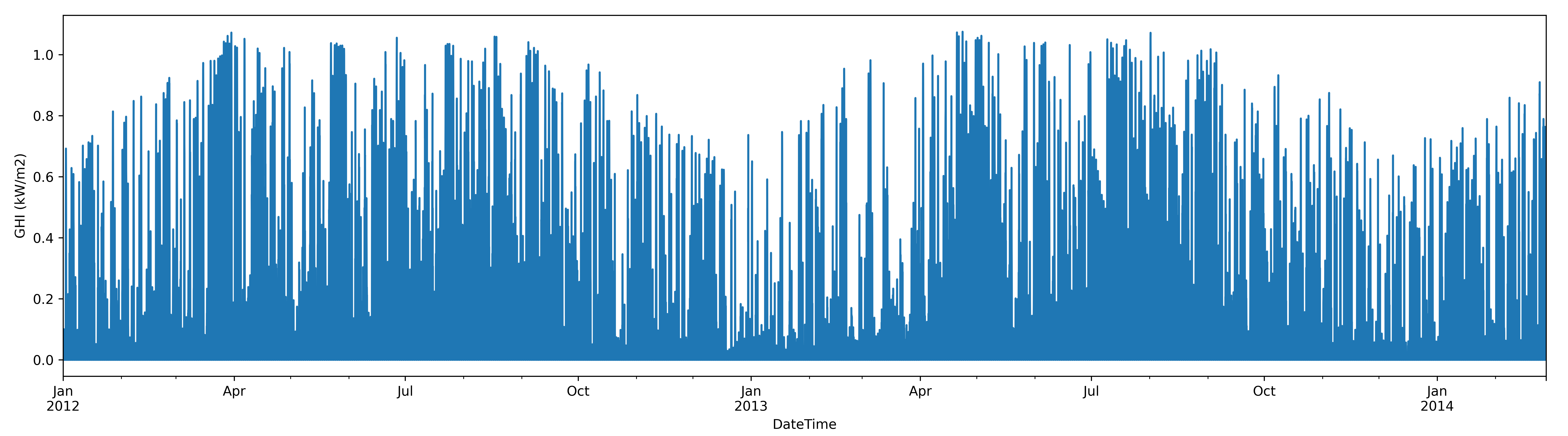}
\caption{Real-time hourly GHI of London city from Jan 2012 to Feb 2014.}
\label{GHI}
\end{figure}

\begin{figure}[]
\centering
\includegraphics[scale=0.035]{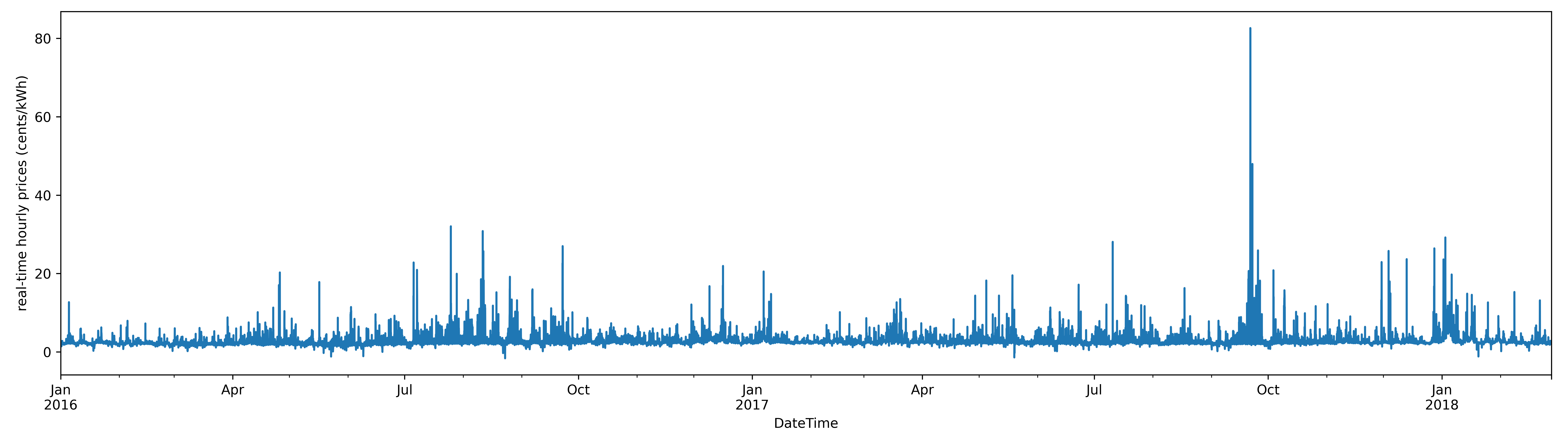}
\caption{Real-time hourly prices of Michigan state from Jan 2016 to Feb 2018.}
\label{RTP}
\end{figure}

In our simulations, real-time hourly solar irradiation of London, UK from Jan 2012 to Feb 2014 shown in Fig. \ref{GHI} is extracted from database of ``Photovoltaic Geographical Information System'' of European Commission \cite{irradiation}. For real-time hourly prices, because we do not have real-time prices of London city, real-time hourly prices of Michigan from Jan 2016 to Feb 2018 obtained from Pecan Street database \cite{prices} are used as shown in Fig. \ref{RTP}. To be specific, in all above datasets, the data from Jan 2012 to Jan 2014 are used to train the neural networks of our proposed strategies and the data of Feb 2014 are used to test and evaluate the performance of our proposed strategies.

Main parameters of ESS, RES and MADDPG algorithm are shown in Table \ref{main_parameters}. In this table, $N_{RES}^{a}$, $N_{RES}^{c}$, $N_{ESS}^{a}$, $N_{ESS}^{c}$ are the capacity of the actor and the critic network of agents RES and ESS, respectively. $lr_a$, and $lr_c$ are the learning rate of the actor network and critic network, respectively.
\begin{table}[ht]
\caption{Main parameters of ESS, RES and MADDPG in our study.}
\centering
\begin{tabular}{|c|c|c|c|}
\hline
$\eta^{ESS}$ &  $0.9$ & $Ch_{rate}$ & $1.0$ \textit{kW} \\
\hline
$Dh_{rate}$ & $1.0$ \textit{kW} & $EL_{max}$ & $10$ \textit{kWh} \\
\hline
$EL0$ &  $0.5$ \textit{kWh} & $EL_{min}$ & $0.5$ \textit{kWh} \\
\hline
$\eta^{RES}$ & $0.9$ & $S$ & $1$ \textit{$m^2$} \\
\hline
$\alpha_{p}$ & $1$ & $\gamma$ & $0.99$ \\
\hline
$N_{RES}^{a}$ & $100:100$ & $N_{RES}^{c}$ & $100:200:200$ \\
\hline
$N_{ESS}^{a}$ & $200:200:200$ & $N_{ESS}^{c}$ & $100:200:200$ \\
\hline
$lr_a$ & $0.001$ & $lr_c$ & $0.0001$ \\
\hline
$\tau$ & $0.001$ & $Episodes$ & $4000$ \\
\hline
$Activation$ & $relu$ & $Optimizer$ & $Adam$ \\
\hline
\end{tabular}
\label{main_parameters}
\end{table}

\subsection{Performance evaluation and comparison of three strategies}
\begin{figure}[]
\centering
\includegraphics[scale=1]{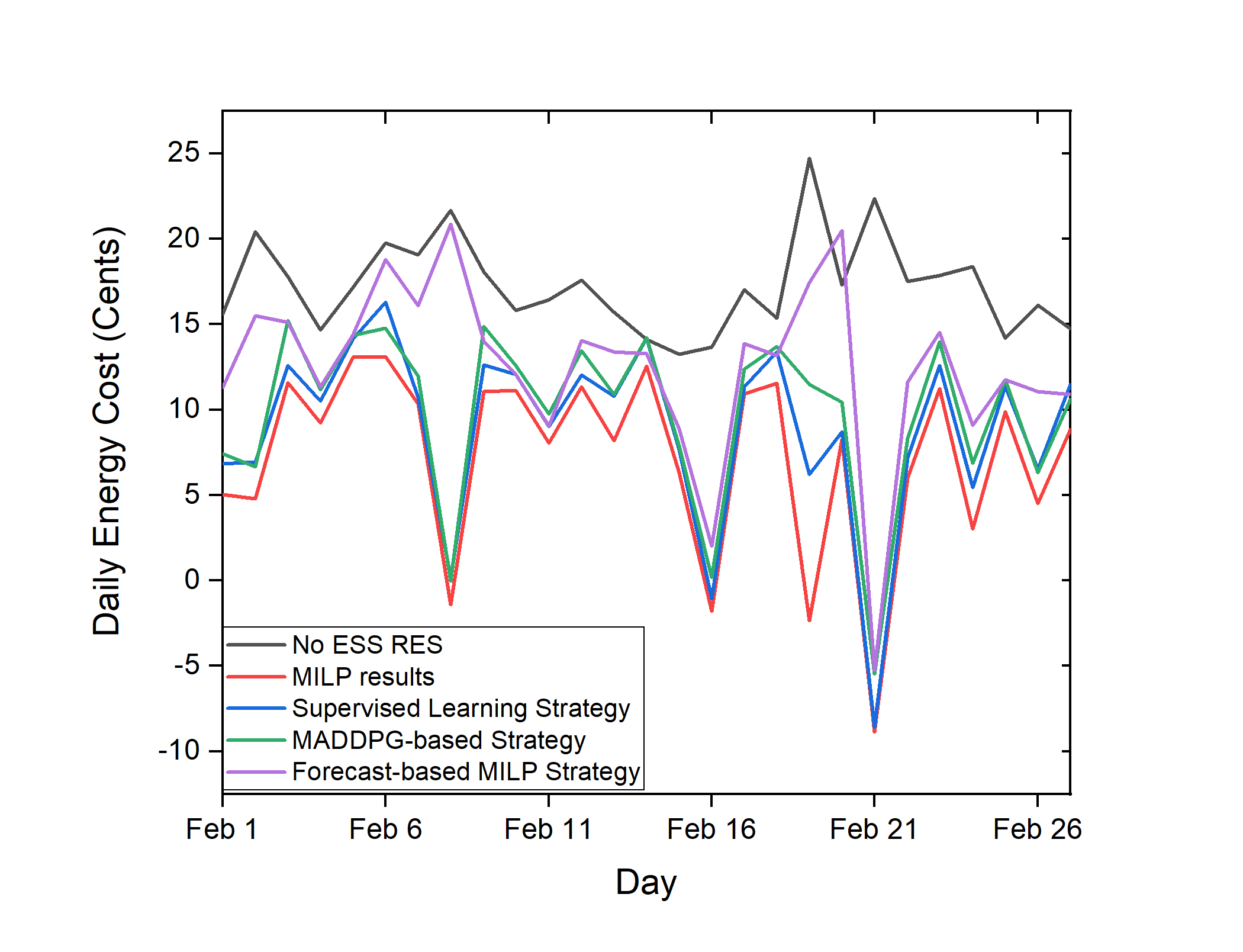}
\caption{The daily energy costs of three strategies at the stable home in Feb 2014. }
\label{stable_result}
\end{figure}

Fig. \ref{stable_result} shows the energy cost achieved by three strategies of each day in Feb 2014 (testing set) at the stable home. The black line is the daily energy cost without ESS and RES whereas the red line is optimal daily energy cost which can be achieved by using MILP solver at the end of each day when ESS and RES is fully utilized and the full information of that day is already known. Clearly, these MILP results (red line) are best daily energy cost we can achieve and this line is a lower bound of our proposed strategies. As shown in this figure, the daily energy cost achieved by MILP-based supervised learning is the closest to the red line whereas the daily energy cost achieved by forecast-based strategy is the farthest away from the red line among three strategies. Hence, the performance of MILP-based supervised learning strategy is best and the performance of forecast-based strategy is worst among three proposed strategies.

\begin{figure}[]
\centering
\includegraphics[scale=1]{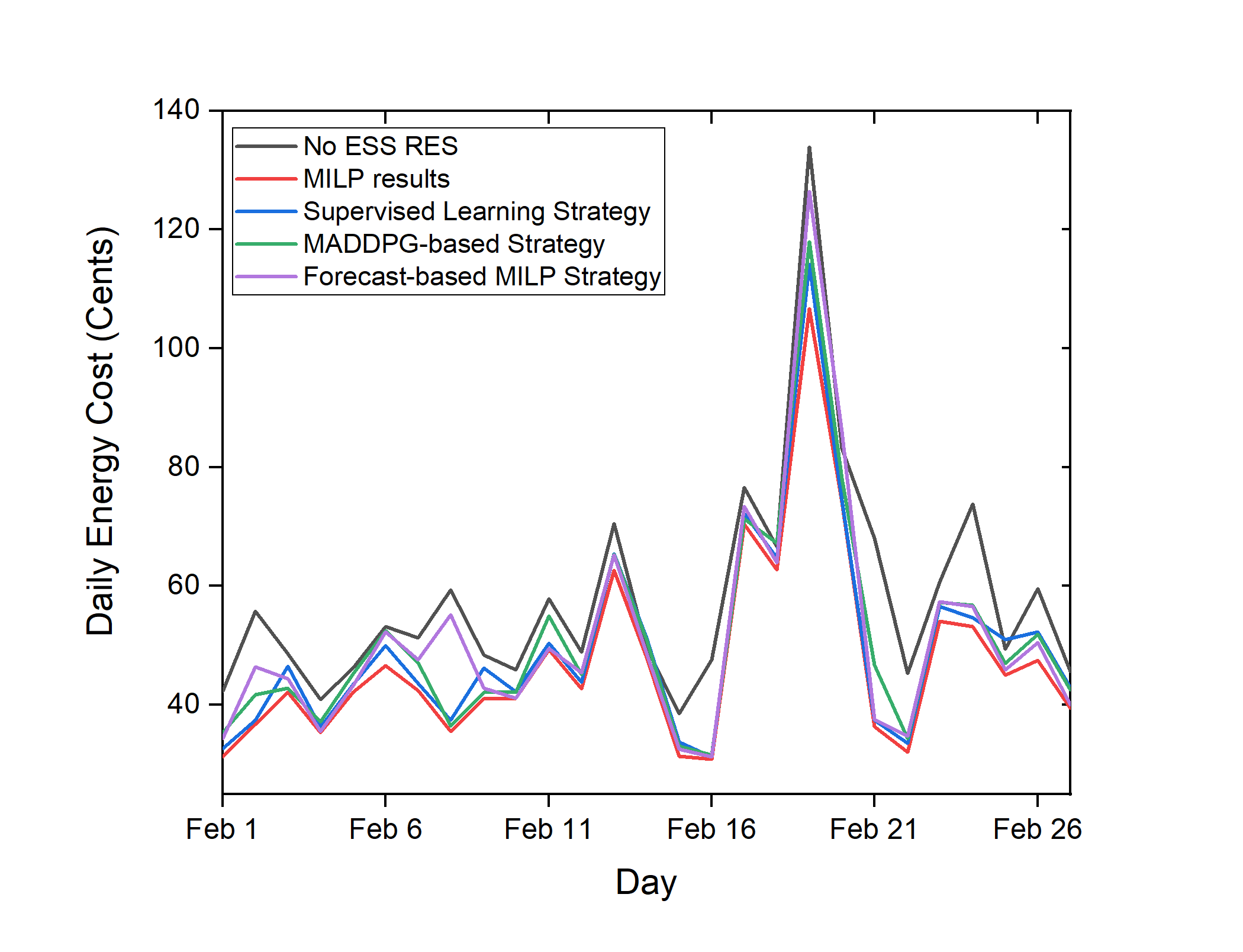}
\caption{The daily energy costs of three strategies at the fluctuating home in Feb 2014. }
\label{fluctuating_result}
\end{figure}

\begin{figure}[]
\centering
\includegraphics[scale=1]{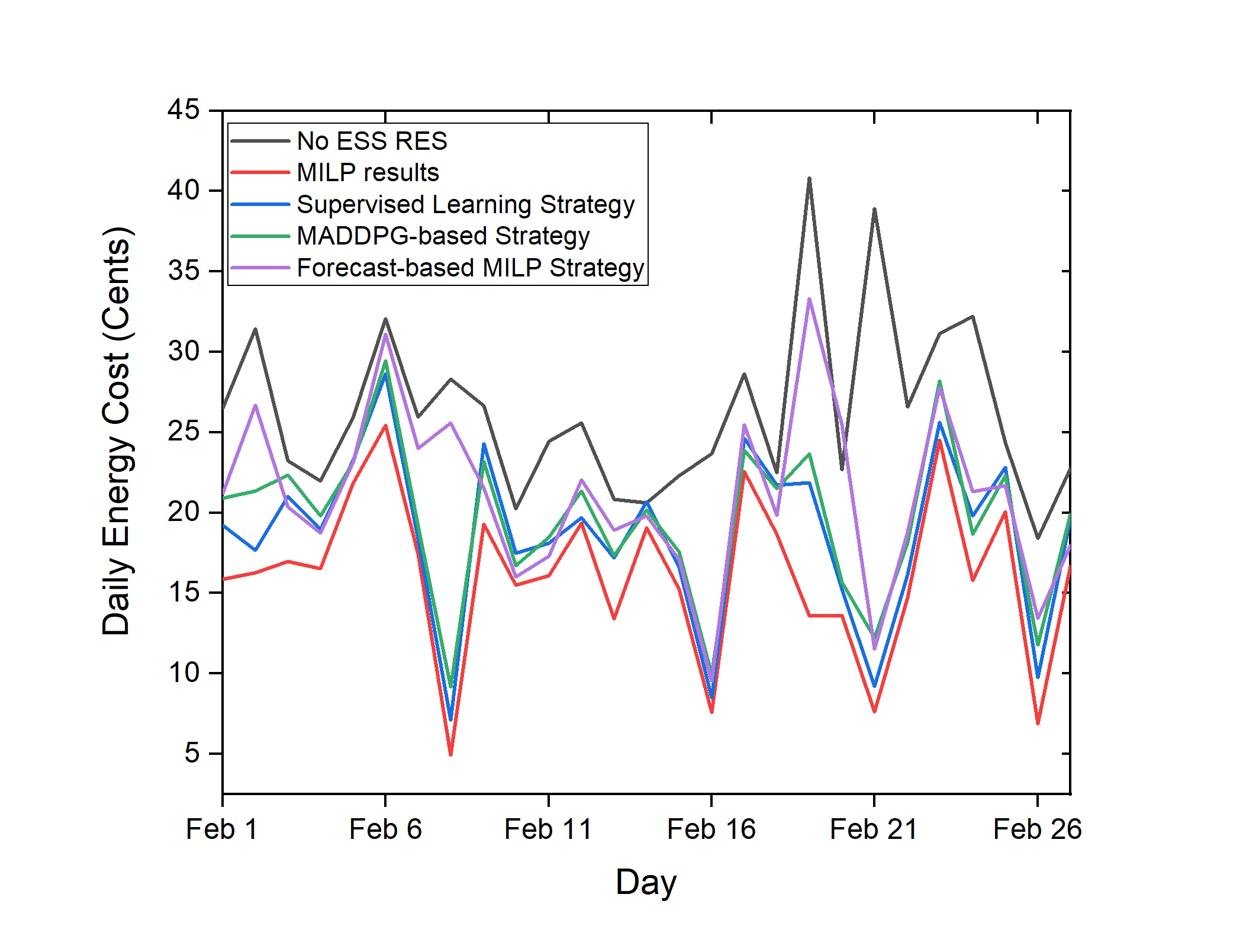}
\caption{The daily energy costs of three strategies at the chaos home in Feb 2014. }
\label{chaos_result}
\end{figure}

Similarly, at the fluctuating home and chaos home, the daily energy cost of MILP-based supervised learning strategy is also closer to the red line than those of other strategies as shown in Fig. \ref{fluctuating_result} and Fig. \ref{chaos_result}, respectively. Hence, the performance of MILP-based supervised learning strategy is also the best among three strategies whereas the performance of forecast-based strategy is also the worst at these homes.

Clearly, one of main objectives of using ESS and RES is to reduce the daily energy cost as much as possible and the cost saving achieved by MILP solver is maximum saving we can achieve. For a better comparison between our strategies, we introduce a new metric shown in (\ref{eff}) to calculate the effectiveness of each proposed strategies in testing set. This metric shows us how much cost saving we can achieve on average by each strategy compared with cost saving by MILP solver in percentage terms.
\begin{equation}
Eff_{MILP}^{strategy}= \frac{\displaystyle\sum_{i=1}^{N} \frac{C_{base}^{i} - C_{strategy}^{i}}{C_{base}^{i}}}{\displaystyle\sum_{i=1}^{N} \frac{C_{base}^{i} - C_{MILP}^{i}}{C_{base}^{i}}} \times 100
\label{eff}
\end{equation}
where $N$ is the number of days in testing set. $C_{base}^{i}$ is the energy cost of the day $i$ without ESS and RES (the black line). $C_{MILP}^{i}$ is the energy cost of the day $i$ which is achieved by using the MILP solver at the end of that day (the red line). $C_{strategy}^{i}$ is the energy cost of the day $i$ which is achieved by applying our strategy.
\begin{figure}[]
\centering
\includegraphics[scale=1]{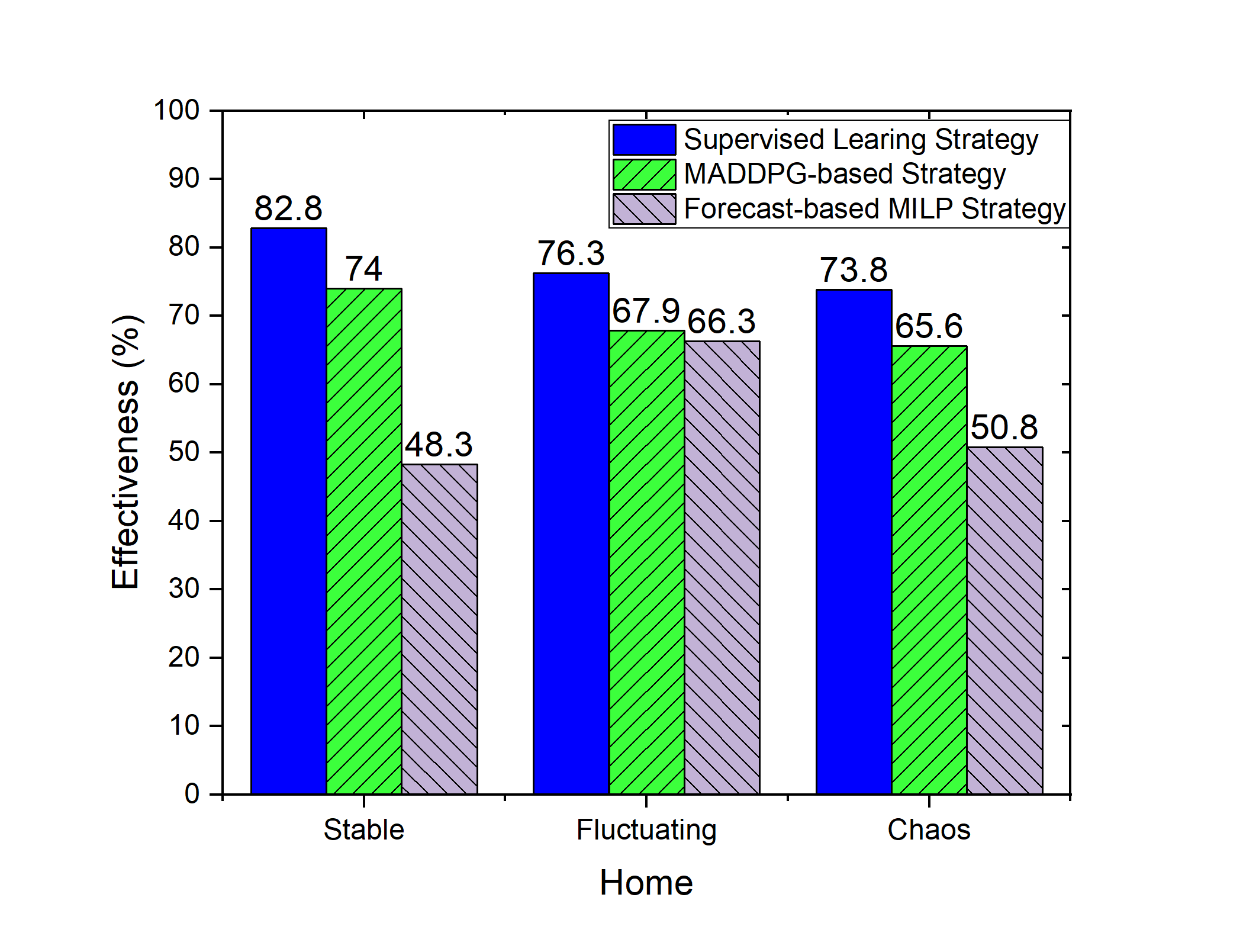}
\caption{The effectiveness of three strategies at three homes in Feb 2014.}
\label{effectiveness}
\end{figure}

\begin{figure}[]
\centering
\includegraphics[scale=1]{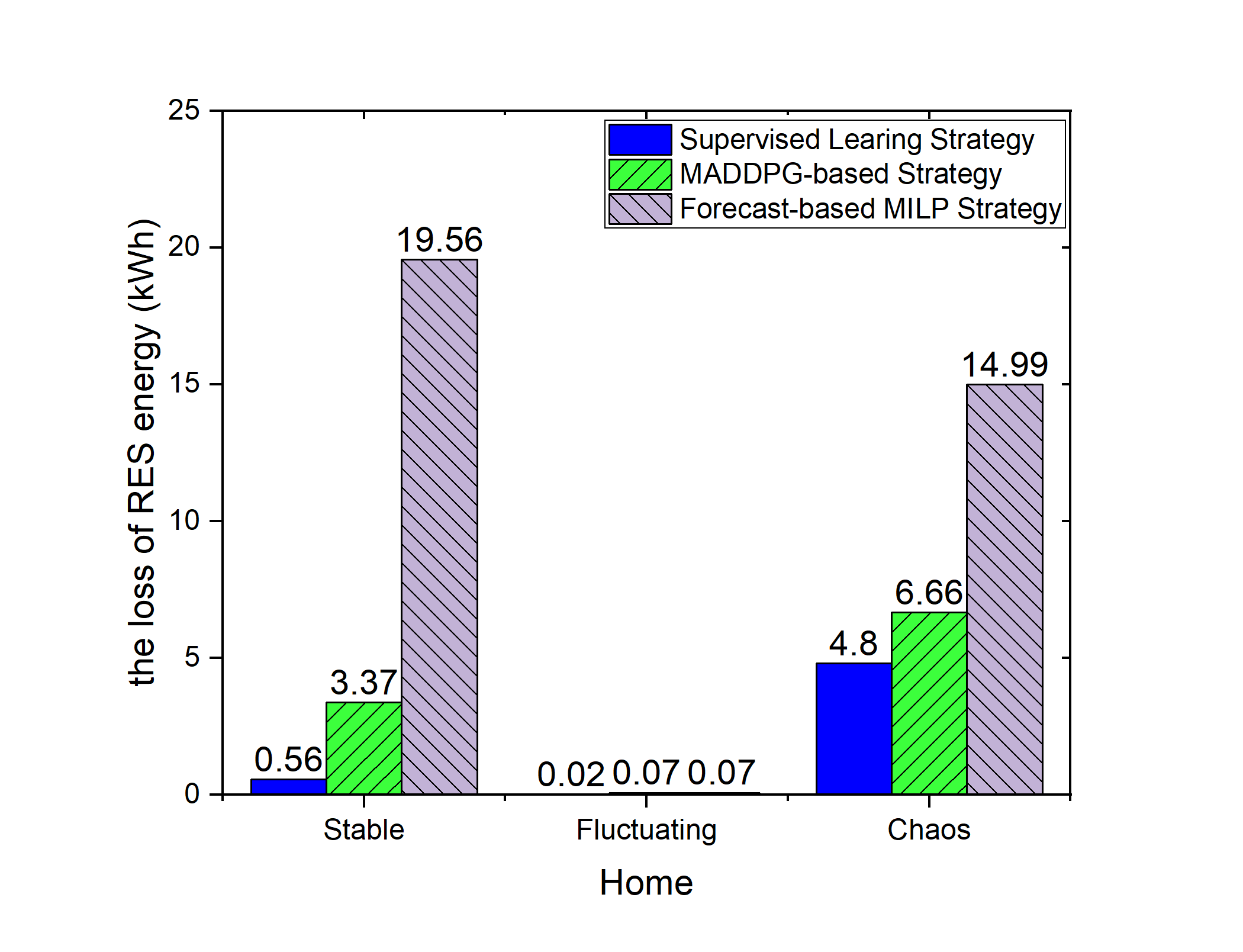}
\caption{The total loss of RES energy at three homes in Feb 2014.}
\label{RES_loss}
\end{figure}

Fig. \ref{effectiveness} shows us the effectiveness of each strategy at three different homes in Feb 2014. As shown in this figure, at the stables home, the average cost saving achieved by MILP-based supervised learning strategy is about $82.8\%$ of the average cost saving achieved by the MILP solver whereas the average cost saving achieved by MADDPG-based strategy and forecast-based strategy are only $74\%$ and $48.3\%$, respectively. At fluctuating and chaos homes, the average cost saving achieved by MILP-based supervised learning is also higher than those of other strategies. These results again confirm that the performance of MILP-based supervised learning strategy is the best and that of forecast-based strategy is the worst among three proposed strategies. We have these results because, in this study, the DNNs of the MILP-based supervised learning approximate the optimal results of MILP solver better than that of MADDPG-based strategy whereas the forecast errors of the $24$-hour predictions worsen the performance of forecast-based strategy. It is worth noting that the average cost saving achieved by forecast-based strategy at fluctuating home is larger than those achieved by forecast-based strategy at other homes because amount of the hourly energy consumption at fluctuating home is the largest one in three kinds of homes. When the controlling of ESS and RES becomes worse in forecast-based strategy, the performance of this strategy in fluctuating home is improved by absorbing a lot of RES energy. As shown in Fig. \ref{RES_loss}, the loss of RES energy in the fluctuating home is smallest among in three kinds of homes at all strategies.

Table \ref{computational} shows the average computational time of MILP-based supervised learning strategy and MADDPG-based strategy at each time slot. As shown in this table, the computational time of MILP-based supervised learning is larger than that of MADDPG-based strategy because the supervised learning needs more DNNs to predict than MADDPG-based strategy at each time slot. Moreover, the DNNs of supervised learning are built by using Keras, a high-level deep learning APIs of Tensorflow whereas the agents of MADDPG-based strategy are implemented by using low-level APIs of Pytorch. However, the computational time of these strategies is very small. In this table, the computational time of forecast-based strategy is not shown and compared because this strategy is a kind of day-ahead strategy while two remaining strategies are hour-ahead strategies.
\begin{table}[]
\caption{The average computational time.}
\centering
\begin{tabular}{|c|c|}
\hline
Strategy &  Average computational time (s) \\
\hline
MILP-based supervised learning &  $0.18$ \\
\hline
MADDPG-based strategy &  $0.001$  \\
\hline
\end{tabular}
\label{computational}
\end{table}

\section{Discussion}
The proposed MILP-based supervised learning does not intervene home appliances as many traditional methods but silently minimizes daily energy cost. It means that residents still keep their behaviors of using appliances and maximize their comfortable lifestyle. However, as shown in Fig. \ref{effectiveness}, the performance of our proposes strategies decreases steadily when resident behavior is getting more and more chaotic. To apply our strategy to smart homes or buildings, residents should have a good behavior of using their home devices or the MAPE of energy consumption prediction for next $1$ hour should be smaller than $21.8\%$.

Although forecast-based strategy, a kind of day-ahead strategy, is the worst strategy among our proposed strategies, it does not mean that this strategy should be replaced. In the meanwhile, hour-ahead and day-ahead strategy should be combined in smart houses. Day-ahead strategy helps residents determine energy demand and easily take part in energy markets. In the meanwhile, hour-ahead strategy helps residents improve the performance of HEMS and achieve maximum comfortable lifestyle. Moreover, the running time of our proposed strategies is very tiny, implying that it is very potential when integrated in the coordinated DR of multiple HEMSs.

\section{Conclusion}
In this study, we proposed a MILP-based supervised learning strategy for hour-ahead DR of a HEMS which learned resident behavior of using appliances and applied its knowledge to minimize daily energy cost. The DNNs of this strategy were trained by using datasets created by a MILP solver with historical data to approximate MILP optimization. After training, at each time slot, these DNNs were used to control ESS and RES with current real-time information of surrounding environment. This strategy was also compared to MADDPG-based hour-ahead strategy and forecast-based day-ahead strategy. Three different case studies were conducted based on resident behavior of using appliances. The case study results demonstrated the effectiveness of MILP-based supervised learning in terms of daily energy cost. The average cost saving of this strategy is achieved up to $73.8\%$ of average cost saving achieved by MILP solver at home in which the MAPE of $1$-hour prediction is only $21.8\%$. This strategy was also the best strategy among three proposed strategies at all case studies.

Future works will focus on the development of a cooperative strategy for group of homes where a resident at each home will buy or sell energy together. This strategy will be possible an extended version of our proposed strategy. Basically, our MILP-based supervised learning strategy can be considered as MILP-based IL strategy in which our DR tries to mimic the MILP solver by using supervised learning technique, a traditional approach of IL. Hence, another way to improve our study is to apply advanced algorithms of IL such as DAgger \cite{ross2011reduction}.


%

\appendices






\ifCLASSOPTIONcaptionsoff
  \newpage
\fi



\bibliographystyle{IEEEtran}
\bibliography{my_references}
%

%




\end{document}